\input{aipcheck}

\documentclass[
    ,final            
  ]
  {aipproc}

\layoutstyle{6x9}

\newcommand{\rfp}{\sqrt{4\pi}}
\newcommand{\freq}{{\rm e}^{i\epsilon t}}
\newcommand{\eal}{\left(\epsilon-\frac{\alpha}{2}\omega\right)} 
\newcommand{\fract}[2]{\textstyle{\frac{#1}{#2}}}

\begin{document}

\title{Magnetic Moments of Baryons with a Heavy Quark\footnote{Talk
presented by HW at MRST 03 (Joe-Fest) Syracuse, NY, May 2003.}}

\author{H. Weigel}{
  address={Institute for Theoretical Physics, T\"ubingen University\\
Auf der Morgenstelle 14, D--72076 T\"ubingen, Germany}
}

\author{S. Scholl}{
  address={Institute for Theoretical Physics, T\"ubingen University\\
Auf der Morgenstelle 14, D--72076 T\"ubingen, Germany}
}

\begin{abstract}
We compute magnetic moments of baryons with a heavy 
quark in the bound state approach for heavy baryons.
In this approach the heavy baryon is considered as a 
heavy meson bound to a light baryon. The latter is
represented as a soliton excitation of light meson
fields. We obtain the magnetic moments by sandwiching
pertinent components of the electromagnetic current
operator between the bound state wave--functions. We 
extract this current operator from the coupling to 
the photon field after extending the action to be 
gauge invariant.
\end{abstract}

\maketitle


\section{Introduction}

This talk is dedicated to Joe Schechter on the occasion of his 
${\rm 65}^{\rm th}$ birthday. Joe has made important contributions 
to the descriptions of hadrons with a heavy quark~\cite{Sch92,Ja95}. 
In particular he has successfully developed a hadron model that is 
valid for finite masses of the heavy quarks but naturally leads into 
the heavy quark effective model as these masses tend towards 
infinity~\cite{Sch93,Sch95heavy,Ha97}.
In this talk we will present an application of this approach.

The basic idea is to consider an effective meson 
model that includes both, light degrees of freedom and
mesons that contain a heavy quark. In this model the
interaction of the light degrees of freedom is governed
by chiral symmetry. On the other hand the heavy quark 
components of the heavy meson field are coupled such that
the Lagrangian reproduces the heavy quark symmetry, {\it i.e.} 
spin and flavor independence of the interactions, as the
masses of the heavy mesons tend to infinity. It is thus 
essential that this part of the model contains both the 
heavy pseudoscalar and heavy vector meson fields.

It is well established that the light sector of such a model may 
contain soliton solutions to the classical equations of 
motion~\cite{Ja87,Me89}. After appropriate quantization these 
solutions may be identified with baryon states as in the Skyrme model 
framework~\cite{Sk61}. Essentially this is a practical realization 
of the large $N_C$ picture for baryons in QCD~\cite{Wi79}. The 
heavy mesons interact with the light mesons according to the roles 
of chiral symmetry. Substituting the soliton for the light meson
fields into that interaction Lagrangian generates a background potential 
for the heavy meson field. This potential has bound state solutions 
that, when combined with the soliton, describe baryons with a 
heavy quark~\cite{Ca85,Ku89,Oh91}. It is then, of course, 
interesting to study properties of such a heavy baryon by 
computing matrix elements of pertinent operators. The most prominent 
operator that comes to mind is the one of the electromagnetic
current. In particular the spatial components
of this operator determine the magnetic moments of the considered
baryon.  We obtain this operator by incorporating the photon field 
such that the total action becomes gauge invariant. The current 
operator is then the coefficient that is linear in the photon field. 

This presentation is based on on--going studies regarding 
electromagnetic properties of baryons with a heavy quark 
in the above described bound state picture. These studies will 
be presented elsewhere~\cite{Sch03}. Therefore the numerical
results presented here should be considered preliminary.
We will commence this talk with a brief a review of the soliton 
picture for the light mesons when vector mesons like $\rho$ and 
$\omega$ are included. Such an extension has proven necessary in many 
aspects, as {\it e.g.} the proton--neutron mass difference~\cite{Ja89}, 
the predicted pion--nucleon phase shifts~\cite{Schw89} or the axial 
singlet charge of the nucleon~\cite{Jo90}. We will then discuss the 
occurance of the bound state and describe the quantization of the 
combined soliton bound state system that leads to the heavy baryon 
state. Finally we will compute the magnetic moments of heavy baryons
as matrix elements of such bound states.

\section{The Model Lagrangian}

In this section we review the classical, {\it i.e.} leading 
order part in the $1/N_C$ expansion of the bound state 
description for the heavy baryons in the soliton picture.

\bigskip
\subsection{Light Mesons}
\smallskip

For the sector of the model describing the light pseudoscalar
and vector mesons we adopt the chirally invariant Lagrangian
discussed in detail in the literature \cite{Ka84,Ja87}.
This Lagrangian can be decomposed into a regular parity part
\begin{eqnarray}
{\cal L}_S=f_\pi^2{\rm tr}\left[p_\mu p^\mu\right]
+\frac{m_\pi^2f_\pi^2}{2}{\rm tr}\left[U+U^{\dag}-2\right]
-\frac{1}{2}{\rm tr}\left[F_{\mu\nu}\left(\rho\right)
F^{\mu\nu}\left(\rho\right)\right]
+m_V^2{\rm tr}\left[R_\mu R^\mu\right]
\label{lagnorm}
\end{eqnarray}
and a part which contains the Levi-Civita tensor,
$\epsilon_{\mu\nu\alpha\beta}$. The action for the latter is 
most conveniently displayed with the help of differential 
forms $p=p_\mu dx^\mu$, etc.:
\begin{eqnarray}
\Gamma_{\epsilon}&&=\frac{2N_c}{15\pi^2}\int Tr(p^5)\cr
&&\quad
+\int Tr\left[\frac{4i}{3}(\gamma_1+\frac{3}{2}\gamma_2)Rp^3
-\frac{g}{2} \gamma_2 F(\rho )(pR-Rp)
-2ig^2 (\gamma_2+2\gamma_3) R^3p\right].
\label{lagannorm}
\end{eqnarray}
In eqs (\ref{lagnorm}) and (\ref{lagannorm}) we have introduced the
abbreviations
\begin{eqnarray}
p_\mu=\fract{i}{2}\left(\xi\partial_\mu\xi^{\dag}-
\xi^{\dag}\partial_\mu\xi\right),\quad
v_\mu=\fract{i}{2}\left(\xi\partial_\mu\xi^{\dag}+
\xi^{\dag}\partial_\mu\xi\right) \quad {\rm and}\quad
R_\mu=\rho_\mu-\fract{1}{g}v_\mu \ .
\label{lightcurrents}
\end{eqnarray}
Here $\xi$ refers to a square root of the chiral field,
{\it i.e.} $U=\xi^2$. Furthermore
$F_{\mu\nu}\left(\rho\right)=\partial_\mu\rho_\nu
-\partial_\nu\rho_\mu-ig\left[\rho_\mu,\rho_\nu\right]$
denotes the field tensor associated with the vector mesons
$\rho$ and $\omega$, which are combined in  
$\rho_\mu=\left(\omega_\mu{\rm 1\!\! I}+\rho_\mu^a\tau^a\right)/2$
when the reduction to two light flavors is made.  The parameters 
$g,\gamma_1,$ etc. can be determined (or at least constrained) 
from the study of decays of the light vector mesons such as 
$\rho\rightarrow2\pi$ or $\omega\rightarrow3\pi$ \cite{Ja87}. 
The complete determination of all parameters, however, also
requires some information for the light baryon sector, see below.

The action for the light degrees of freedom 
($\int {\cal L}_S+\Gamma_{\epsilon}$) contains static soliton
solutions. The appropriate {\it ans\"atze} are~\cite{Ja87}
\begin{eqnarray}
\xi(\vec{r\,})={\rm exp}\left(\frac{i}{2}
\hat{r}\cdot\vec{\tau}\,F(r)\right),
\quad
\omega_0(\vec{r\,})=\frac{\omega(r)}{g}
\quad {\rm and}\quad
\rho_{i,a}(\vec{r\,})=
\frac{G(r)}{gr}\epsilon_{ija}\hat r_j
\label{lightan}
\end{eqnarray}
while all other field components vanish. The resulting non--linear 
Euler--Lagrange equations for the radial functions $F(r),\omega(r)$ 
and $G(r)$ are solved numerically subject to the boundary conditions
$F(0)=-\pi$, $\omega^\prime(0)=0$ and $G(0)=-2$ while all fields vanish
at radial infinity \cite{Ja87}. These boundary conditions are needed 
to obtain a consistent baryon number one configuration.

\bigskip
\subsection{The Relativistic Model for the Heavy Mesons}
\smallskip

In this subsection we present the relativistic Lagrangian, which 
describes the coupling between the light and heavy mesons \cite{Sch93}
\begin{eqnarray}
{\cal L}_H&=&D_\mu P\left(D^\mu P\right)^{\dag}
-\frac{1}{2}Q_{\mu\nu}\left(Q^{\mu\nu}\right)^{\dag}
-M^2PP^{\dag}+M^{*2}Q_\mu Q^{\mu{\dag}}
\cr &&
+2iMd\left(Pp_\mu Q^{\mu{\dag}}-Q_\mu p^\mu P^{\dag}\right)
-\frac{d}{2}\epsilon^{\alpha\beta\mu\nu}
\left[Q_{\nu\alpha}p_\mu Q_\beta^{\dag}+
Q_\beta p_\mu \left(Q_{\nu\alpha}\right)^{\dag}\right]
\cr &&
-\frac{2\sqrt{2}icM}{m_V}\Bigg\{
2Q_\mu F^{\mu\nu}\left(\rho\right)Q_\nu^{\dag}
 \cr &&  \hspace{2cm}
-\frac{i}{M}\epsilon^{\alpha\beta\mu\nu}\left[
D_\beta PF_{\mu\nu}\left(\rho\right)Q_\alpha^{\dag}
+Q_\alpha F_{\mu\nu}\left(\rho\right)\left(D_\beta P\right)^{\dag}
\right]\Bigg\}.
\label{lheavy}
\end{eqnarray}
Here we have allowed the mass $M$ of the heavy pseudoscalar meson
$P$ to differ from the mass $M^*$ of the heavy vector meson $Q_{\mu}$.
Note that the heavy meson fields are conventionally defined as {\it row}
vectors in isospin space with $P=(P^\dagger)^\dagger\,,
D_\mu P= (D_\mu P^\dagger)^\dagger $ etc.. The covariant 
derivative introduces the additional parameter $\alpha$:
\begin{eqnarray}
D_\mu P^{\dag}&=&\left(\partial_\mu-i\alpha g \rho_\mu
-i\left(1-\alpha\right)v_\mu\right)P^{\dag}
=\left(\partial_\mu-iv_\mu-ig\alpha R_\mu\right)P^{\dag}\ ,
\label{covderp} \\
D_\mu Q_\nu^{\dag}&=&
\left(\partial_\mu-iv_\mu-ig\alpha R_\mu\right)Q_\nu^{\dag} \ .
\label{covderq}
\end{eqnarray}
The covariant field tensor of the heavy vector meson is then 
defined as
\begin{eqnarray}
\left(Q_{\mu\nu}\right)^{\dag}=
D_\mu Q_\nu^{\dag}-D_\nu Q_\mu^{\dag}.
\label{heavyft}
\end{eqnarray}
The coupling constants $d,c$ and $\alpha$, which appear in the 
Lagrangian (\ref{lheavy}), have still not been very accurately 
determined. In particular there is no direct experimental evidence 
for the value of $\alpha$, which would be unity if a possible 
definition of light vector meson dominance for the electromagnetic 
form factors of the heavy mesons was adopted~\cite{Ja95}. The
other parameters in (\ref{lheavy}) will be taken~\cite{Ja95} to be:
\begin{eqnarray}
d&=&0.53\ , \quad c=1.60\ ;
\nonumber \\
M&=&1865{\rm MeV}\ , \quad M^*=2007\,{\rm MeV}\ , \qquad {\rm D-meson}\ ;
\nonumber \\
M&=&5279{\rm MeV}\ , \quad M^*=5325\,{\rm MeV}\ , \qquad {\rm B-meson}.
\label{heavypara}
\end{eqnarray}

It should be stressed that the assumption of infinitely large
masses for the heavy mesons has not been made in (\ref{lheavy}).
However, a model Lagrangian which was only required to exhibit the 
Lorentz and chiral invariances would be more general than the 
relativistic Lagrangian (\ref{lheavy}). Rather the 
coefficients of the various Lorentz and chirally invariant pieces 
in the relativistic Lagrangian (\ref{lheavy}) have precisely been 
arranged to become spin--flavor symmetric in the heavy quark 
limit $M,M^\ast\to\infty$~\cite{Sch93}.

\bigskip
\subsection{Bound States}
\smallskip

Here we briefly review the origin of bound states in the S-- and 
P--wave heavy meson channels. These orbital angular momentum quantum 
numbers refer to those of the pseudoscalar component ($P^{\dag}$) of 
the heavy meson multiplet ($P^{\dag},Q_\mu^{\dag}$). More details
are provided in refs.~\cite{Sch95heavy}.

For the P--wave channel the appropriate {\it ansatz} reads
\begin{eqnarray}
P^{\dag}&=&\frac{1}{\rfp}\Phi(r)\hat{r}\cdot\vec{\tau}
\rho \freq \ ,  \qquad
Q^{\dag}_0=\frac{1}{\rfp}\Psi_0(r)\rho \freq \ , 
\cr
Q^{\dag}_i&=&\frac{1}{\rfp}\left[i\Psi_1(r){\hat r}_i
+\frac{1}{2}\Psi_2(r)\epsilon_{ijk}{\hat r}_j\tau_k\right]
\rho \freq \ .
\label{pansatz}
\end{eqnarray}
Note that here $\rho$ refers to a properly normalized spinor 
which describes the isospin of the heavy meson multiplet.
There exist also S--wave bound states 
($P^{\dag}=\frac{1}{\rfp}\Phi(r)\rho \freq,\ldots$) which are, 
however, not relevant for the discussion of the magnetic 
moments. For more details see ref.~\cite{Sch95heavy}.

We substitute the soliton background~(\ref{lightan}) for the 
light fields into the Lagrangian~(\ref{lheavy}) and
find the equations of motion for the radial functions in 
the ansatz~(\ref{pansatz}). Due to the soliton background 
there are solutions to these equations with
$|\epsilon|<M$. These are the bound states we are looking for.

\begin{figure}
\centerline{
\includegraphics[height=.28\textheight,width=.4\textwidth,angle=270]{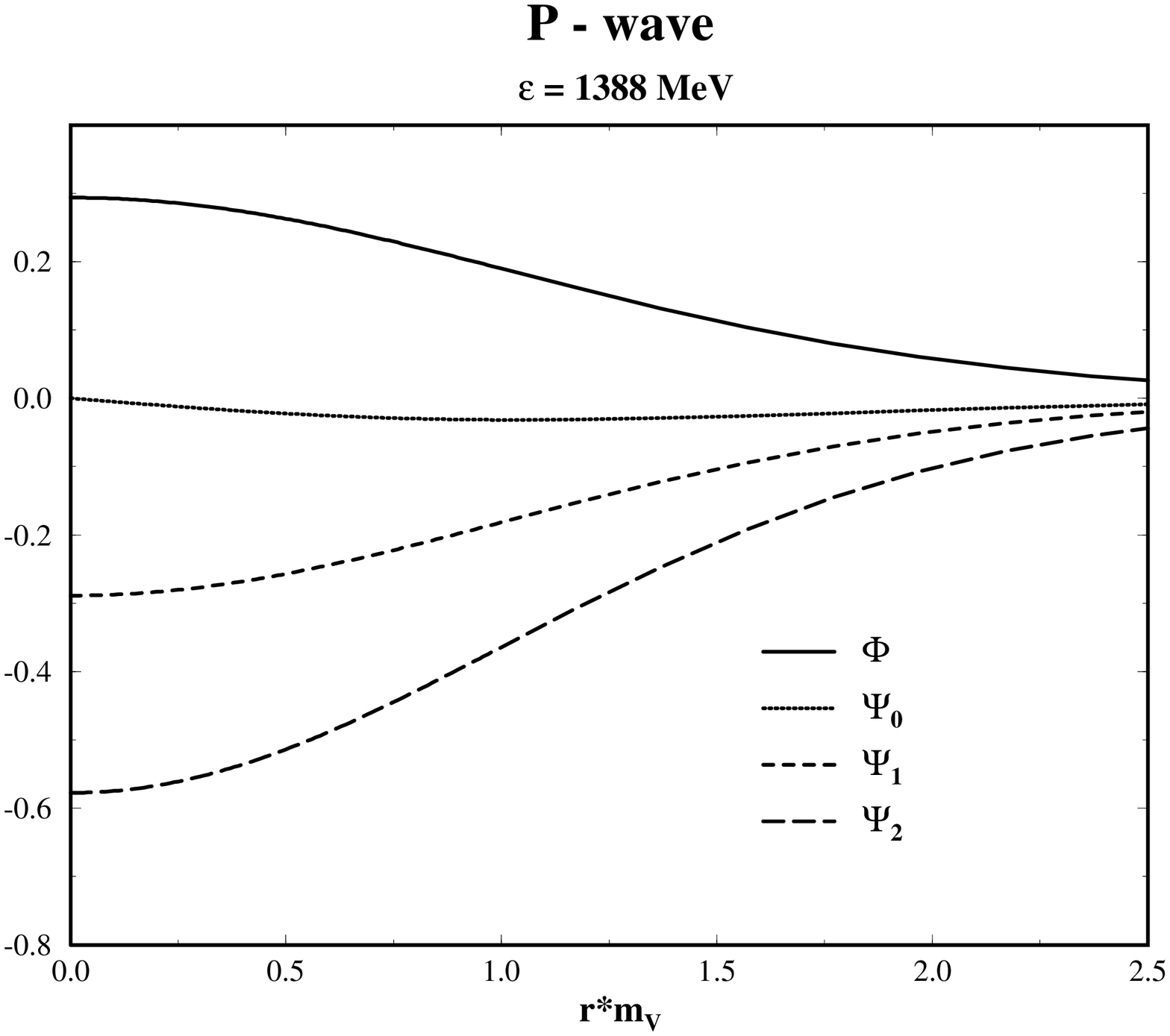}
\hspace{1.2cm}
\includegraphics[height=.28\textheight,width=.4\textwidth,angle=270]{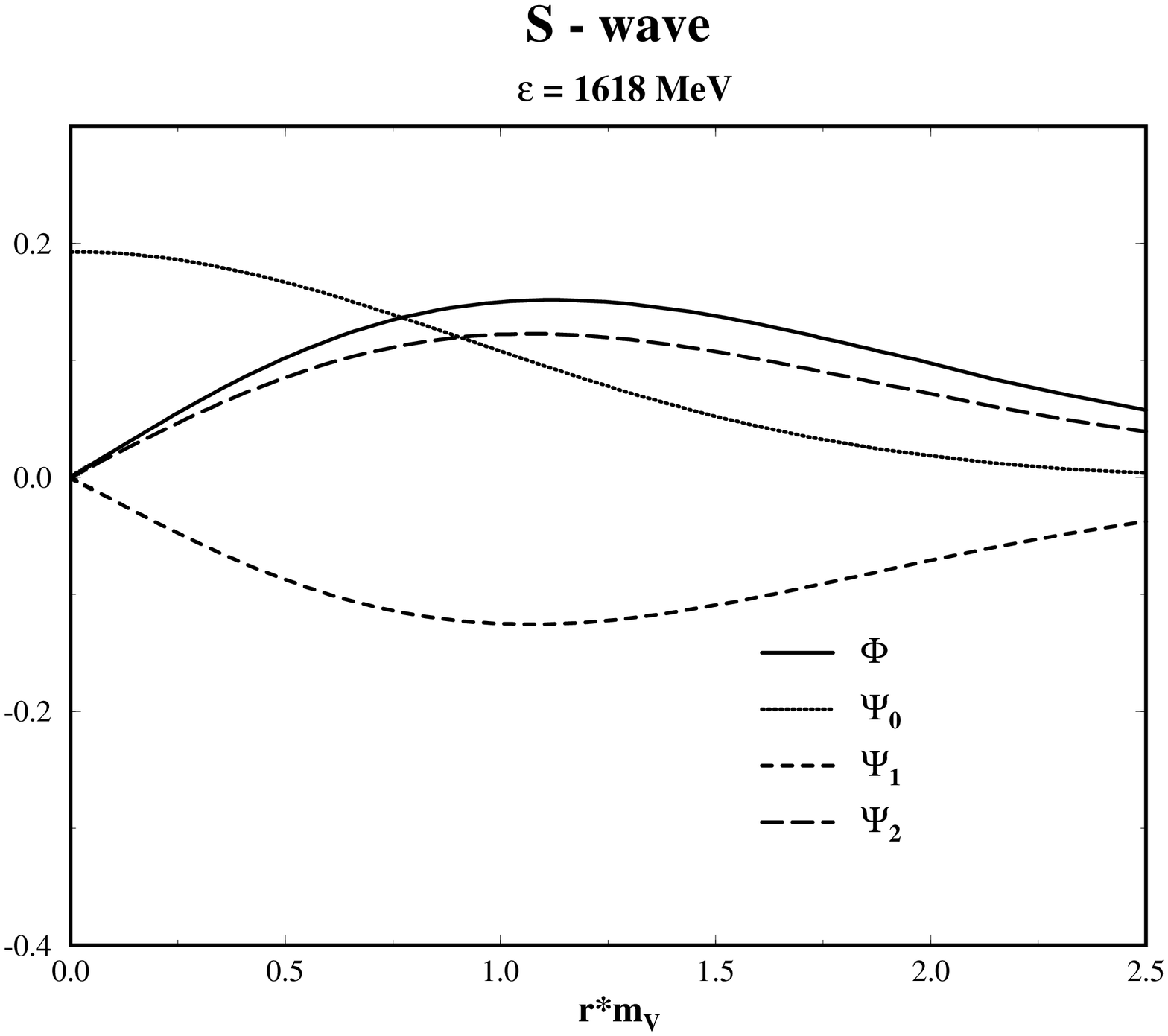}}
\caption{\label{fig_1}\baselineskip16pt
The profile functions for the bound state wave--functions in the 
P--wave (left panel) and S--wave (right panel) channels. These 
functions are measured in units of $m_V=773{\rm MeV}$. See text for 
the specification of the remaining parameters. Figures adopted from
ref.~\protect\cite{Sch95heavy}.}
\end{figure}

As the relativistic Lagrangian (\ref{lheavy}) is bilinear 
in the heavy meson fields the resulting equations of motion 
are linear. Hence the overall magnitude of the solution 
is not fixed by the equation of motion. Nevertheless, the
equations of motion for the heavy meson fields allow us 
to extract a metric for a scalar product between different 
bound states. In particular its diagonal elements serve 
to properly normalize the bound state wave--functions. The Lagrange 
function which results from substituting the {\it ans\"atze} 
(\ref{lightan}) and (\ref{pansatz}) may generally be written as 
\begin{eqnarray}
L=- M_{\rm cl}\left[F,G,\omega\right]+
I_\epsilon\left[F,G,\omega;\Phi,\Psi_0,\Psi_1,\Psi_2\right]
\rho^{\dag}(\epsilon)\rho(\epsilon) \ .
\label{laggen}
\end{eqnarray}
Here $M_{\rm cl}$ denotes the soliton mass \cite{Ja87} whose 
minimum determines the light meson profiles $F,G$ and $\omega$.
The explicit expressions for the functionals $I_\epsilon$ are given 
in ref.~\cite{Sch95heavy}. The subscript refers to the explicit 
dependence on the energy eigenvalues. Upon canonical quantization 
the Fourier amplitudes $\rho(\epsilon)$ and $\rho^{\dag}(\epsilon)$ 
are respectively elevated to annihilation and creation operators for 
a heavy meson bound state with the energy eigenvalue~$\epsilon$. 
Demanding that each occupation of the bound state adds the amount 
$|\epsilon|$ to the total energy yields the normalization 
condition
\begin{eqnarray}
\Big|\frac{\partial}{\partial\epsilon}
I_\epsilon\left[\Phi,\Psi_0,\Psi_1,\Psi_2\right]\Big|=1
\label{normg}
\end{eqnarray}
in addition to the canonical commutation relation
$[\rho_i(\epsilon),\rho^{\dag}_j(\epsilon^\prime)]=
\delta_{ij}\delta_{\epsilon,\epsilon^\prime}$. Note that
for bound states the energy eigenvalues are discretized. For 
the P--wave channel we obtain the normalization condition
{\small
\begin{eqnarray}
&&\Bigg| \int dr r^2 \Bigg\{2\eal\Phi^2
-2\left[\Psi_0^\prime-\eal\Psi_1\right]\Psi_1
+\left[\frac{1}{r}R_\alpha\Psi_0+\eal\Psi_2\right]\Psi_2
\nonumber \\* && \hspace{1cm}
-d\left[\frac{2}{r}{\rm sin}F\Psi_1
-\frac{1}{2}F^\prime\Psi_2\right]\Psi_2
+\frac{4\sqrt{2}c}{gm_V}\frac{1}{r^2}
\left[G\left(G+2\right)\Psi_1-G^\prime r\Psi_2\right]\Phi
\Bigg\} \Bigg|=1 \qquad
\label{normp}
\end{eqnarray}
}~\hskip-0.1cm
from eq (\ref{normg}). For convenience we have employed the abbreviation
$R_\alpha={\rm cos}F-1+\alpha\left(1+G-{\rm cos}F\right)$. The analogous
condition for the bound state wave function in the S--wave channel is
given in ref.~\cite{Sch95heavy}.
The radial profiles associated with these normalizations
are displayed in figure \ref{fig_1} for the choice $\alpha=-0.3$.
The parameters in the relativistic Lagrangian (\ref{lheavy}) have 
been set to the charm sector (\ref{heavypara}). The parameters 
entering the light meson Lagrangian (\ref{lagnorm},\ref{lagannorm}) 
are given in eq (\ref{lightpara}).

\section{Cranking the Bound Heavy Meson State}

It can easily be verified that the field configurations for both 
the light mesons (\ref{lightan}) and the heavy mesons 
configurations~(\ref{pansatz}) obtained above are neither eigenfunctions 
of the spin-- nor the isospin generators. Here we will construct
such eigenstates from the soliton~--~bound state system.

\subsection{Collective Coordinates and Their Quantization}

In order to generate states
which correspond to physical baryons a cranking procedure is employed.
In the first step collective coordinates, which parameterize the
(iso--) spin orientation of the meson configuration, are introduced
via
\begin{eqnarray}
\xi\longrightarrow A(t)\xi A^{\dag}(t)
\quad {\rm and} \quad 
\rho_\mu\longrightarrow A(t) \rho_\mu A^{\dag}(t) \ .
\label{colrot}
\end{eqnarray}
The time--dependence of the collective rotations is 
measured by angular velocities~$\vec{\Omega}$
\begin{eqnarray}
A^{\dag}(t)\frac{d}{dt} A(t)=\frac{i}{2}
\vec{\tau}\cdot\vec{\Omega} \ .
\end{eqnarray}
In addition to the collective
rotation of the soliton configuration, field components are induced
that vanish classically. For the light vector mesons these 
are~\cite{Me89} 
\begin{eqnarray}
\omega_i=\frac{2}{r}\varphi(r)\epsilon_{ijk}\Omega_j{\hat r}_k
\quad {\rm and}\quad
\rho_0^k=\xi_1(r)\Omega_k+\xi_2(r)\hat{r}\cdot
\vec{\Omega\,}{\hat r}_k \ .
\label{induced}
\end{eqnarray}
The light meson Lagrangian then contains a term which is quadratic in 
the angular velocities. The constant of proportionality defines 
the moment of inertia 
$\alpha^2\left[F,G,\omega;\xi_1,\xi_2,\varphi\right]$. The 
radial functions $\varphi(r)$, $\xi_1(r)$ and $\xi_2(r)$ are 
obtained from a variational approach to $\alpha^2$~\cite{Me89}.
Here it is worth mentioning that $\alpha^2$ is of order $N_C$.

In analogy to eq (\ref{colrot}) the heavy meson fields also 
need to be rotated in isospin space,
\begin{eqnarray}
P^{\dag}\longrightarrow A(t)P^{\dag}
\quad {\rm and} \quad
Q^{\dag}_\mu\longrightarrow A(t)Q^{\dag}_\mu \ .
\label{rotheavy}
\end{eqnarray}
Substituting the collectively rotating configurations 
into the total Lagrangian finally yields (the subscript 
$\ell=0,1$ refers to S and P wave bound states)
\begin{eqnarray}
L_\ell=-M_{\rm cl}+I_\epsilon^{(\ell)}\rho^{\dag}\rho
+\frac{1}{2}\alpha^2 \vec{\Omega\,}^2
+\frac{1}{2}\chi_\ell \rho^{\dag}\vec{\Omega}\cdot
\vec{\tau}\rho \ .
\label{collp}
\end{eqnarray}
The quantity 
$I_\epsilon^{(\ell)}$ may be extracted from ref.~\cite{Sch95heavy}.
It has already been employed to obtain the bound state profiles 
functions, {\it e.g.} eq.~(\ref{pansatz}). The new quantity is the hyperfine 
parameter $\chi_\ell$ whose explicit expression is also displayed 
in ref.~\cite{Sch95heavy}.

Once the Lagrangian~(\ref{collp}) for the coupling of the collective
coordinates, $A$ to the creation and annihilation operators, $\rho$
of the bound is found, its quantization proceeds along the bound
state approach to the Skyrmion~\cite{Ca85,Ku89,Oh91,We93}. In 
doing so, we have to construct Noether charges for spin and 
flavor. For this purpose we first have to consider the variation 
of the fields under infinitesimal symmetry transformations.
For the isospin transformation we observe
\begin{eqnarray}
\left[\phi,i\frac{\tau_i}{2}\right]
=-D_{ij}(A)\frac{\partial{\dot \phi}}{\partial \Omega_j} 
+ \ldots \ .
\label{isorot}
\end{eqnarray}
Here $\phi$ refers to any of the iso--rotating meson fields and 
the ellipses represent terms, which are subleading in $1/N_C$, as 
{\it e.g.} time derivatives of the angular velocities which
might arise from eq.~(\ref{induced}). Furthermore
$D_{ij}(A)=(1/2)\ {\rm tr}(\tau_i A \tau_j A^{\dag})$ denotes the 
adjoint representation of the collective rotations $A$. From 
eq (\ref{isorot}) we conclude that the total isospin is related 
to the derivative of the Lagrange function with respect to the 
angular velocities
\begin{eqnarray}
I_i=-D_{ij}(A)\frac{\partial L_\ell}{\partial \Omega_j} \ .
\label{isospin}
\end{eqnarray} 
Next we note that the total spin operator $\vec{J}$
contains the grand spin operator $\vec{G}$
\begin{eqnarray}
G_i=J_i+D_{ij}^{-1}(A)I_j = J_i - J^{\rm sol}_i\ .
\label{gspin1}
\end{eqnarray}
with
$\vec{J\,}^{\rm sol}=
\partial L_\ell / \partial \vec{\Omega}$. As a consequence 
of the relation (\ref{isospin}) we have
$(\vec{J}^{\rm sol})^2=\vec{I\,}^2=I(I+1)$.
By construction the light meson fields do not contribute 
to $\vec{G}$. Even more importantly and as has been 
noted before, the pieces of the heavy meson wave--functions 
(\ref{rotheavy}), which are located between the collective 
coordinates $A$ and the spinor $\rho$, have zero grand 
spin too. With the normalization condition (\ref{normp}) one 
then finds
\begin{eqnarray}
\vec{G}=-\rho^{\dag}
\frac{\vec{\tau}}{2}\rho \ .
\label{gspin2}
\end{eqnarray}
This relates the operator 
multiplying the hyperfine parameter in the collective Lagrangian
(\ref{collp}) to the spin and isospin operators. The collective 
piece of the Hamiltonian is obtained from the Legendre transformation
\begin{eqnarray}
H_\ell^{\rm coll}=\vec{\Omega\,}
\cdot\vec{J\,}^{\rm sol}-L_\ell^{\rm coll}
=\frac{1}{2\alpha^2}\left[\vec{J\,}^{\rm sol} 
+\chi_\ell \vec{G\,}\right]^2\ ,
\label{collham}
\end{eqnarray}
Here $L_\ell^{\rm coll}$ refers to the $\vec{\Omega}$
dependent terms in eq (\ref{collp}). Finally the mass formula for 
a baryon with a single heavy quark becomes
\begin{eqnarray}
M_\ell=M_{\rm cl}+|\epsilon_\ell|
+\frac{1}{2\alpha^2}\left[\chi_\ell J(J+1)+(1-\chi_\ell)I(I+1)\right] \ ,
\label{massp}
\end{eqnarray}
where contributions of ${\cal O}(\chi_\ell^2)$, which apparently  
are quartic in the heavy meson wave--function, have been omitted 
for consistency because terms of that order have been excluded from 
the very beginning. 

{}From eq (\ref{massp}) we recognize that the spin degeneracy 
between baryons containing a heavy quark vanishes in the heavy quark 
limit because $\chi_P$ approaches zero. Of course, this result is a 
direct consequence of the spin--flavor symmetry and would not have 
come out in case the various Lorentz and chirally invariant terms in 
eq (\ref{lheavy}) had been chosen arbitrarily.

The parameters in the light sector, eqs.~(\ref{lagnorm},\ref{lagannorm}) 
cannot completely be determined from properties of the corresponding 
mesons. The remaining (limited) parameter space is, however, more than 
fully constrained by a best fit to the mass differences of the low--lying
$\frac{1}{2}^+$ and $\frac{3}{2}^+$ baryons in the light sector.
This yields:
\begin{eqnarray}
g=5.57,\quad && m_V=773\,{\rm MeV}
\nonumber \\
\gamma_1=0.3, \quad \gamma_2&=&1.8, \quad \gamma_3=1.2 \ .
\label{lightpara}
\end{eqnarray}
The resulting mass differences for the light baryons all agree within
about 10\% \cite{Pa91} with experiment. The corresponding moment of inertia 
is $\alpha^2=5.00{\rm GeV}^{-1}$. The parameters in the heavy baryon 
mass formula~(\ref{massp}) are listed in table~\ref{tab_1}. The resulting 
masses are displayed in table~\ref{tab_2}.
\begin{table}
\caption{\label{tab_1}
The parameters of the mass formula~(\protect\ref{massp}) 
for $\alpha=0.3$. Here $\omega=M-|\epsilon|$ denotes the binding 
energy. The subscripts $S$ and $P$ refer to the 
bound states in the $S$ and $P$ wave channels,
respectively.~\bigskip}
\begin{tabular}{l|cccc}
&$\omega_P$ & $\chi_P$ & $\omega_S$ & $\chi_S$ \\
\hline
D-meson: & 478MeV & 0.114  & 247MeV & 0.205 \\
B-meson: & 713MeV & 0.0045 & 573MeV & 0.0052
\end{tabular}
\end{table}
\begin{table}
\caption{\label{tab_2}
Heavy baryons mass differences with respect to $\Lambda_c$ or 
$\Lambda_b$ for $\alpha=0.3$. Primes indicate negative parity
baryons, {\it i.e.} S--wave bound states. All energies are in 
MeV.~\bigskip}
\begin{tabular}{l | c c c c c | c | c}
& $\Sigma_c$ & $\Sigma_c^*$ & $\Lambda_c^\prime$ &
$\Sigma_c^\prime$ & $\Sigma_c^{\prime *}$ & $N$ &$\Lambda_b$\cr
\hline
$M(B)-M(\Lambda_c)$ & 177 & 211 & 238 & 397 & 
458 & -1321 & 3174 \cr
empir. & 168 & 233 & 308 & ? & ? & -1345 & 3356$\pm$50 \cr
\hline
& $\Sigma_b$ & $\Sigma_b^*$ & $\Lambda_b^\prime$ &
$\Sigma_b^\prime$ & $\Sigma_b^{\prime *}$ & $N$\cr
\hline
$M(B)-M(\Lambda_b)$ & 191 & 205 & 161 & 351 &
367 & -4494& 0\cr
empir. & ? & ? & ? & ? & ? & -4701$\pm$50 & 
\end{tabular}
\end{table}
We observe that the model represents a sensible picture 
of the spectrum of baryons with a heavy quark. This even more
motivates the investigation of their properties, as {\it e.g.}
the magnetic moments.

\section{Electro--magnetic Current and Magnetic Moments}

In this part of the talk we would first like to describe the 
extraction of an electromagnetic current operator for
bound state configuration. We then compute the magnetic moments
by sandwiching that operator
between the heavy baryon states, 
\begin{equation}
B(I,I_3;J,J_3)={\cal N}
\sum_{J^{\rm sol}_3,G_3}
C^{JJ_3}_{IJ^{\rm sol}_3,\fract{1}{2}G_3}
D_{I^3,-J^{\rm sol}_3}^{I=J^{\rm sol}}(A)|\fract{1}{2},G_3\rangle \
\label{hbstate}
\end{equation}
that are obtained by coupling the collective coordinate
wave--function (represented by the Wigner $D$--function, $D(A)$) 
together with bound state wave--function to total spin $J$ according 
to eq.~(\ref{gspin1}). The prefactor ${\cal N}$ is a suitably 
chosen normalization constant.

First we introduce the photon field 
$\cal{A_\mu}$ in close analogy to refs.~\cite{Ja87,Me89} for 
the light sector and ref.~\cite{Ja95} for the sector of the 
heavy mesons. The aim is to construct a gauge invariant 
Lagrangian ${\cal L}(\xi,\rho,P,Q;\cal{A})$ and to extract
the current operator as the term that is linear in the 
photon field: 
$J_\mu^{\rm e.m.}=\partial {\cal L}/ \partial{\cal A}^\mu\,
\big|_{{\cal A}^\mu=0}$.
For the light sector, this exercise has been done in 
refs.~\cite{Ja87,Me89}, while ref.~\cite{Ja95} provides 
the electromagnetic generalization of the covariant 
derivatives~(\ref{covderp},\ref{covderq}). Here we also need 
to gauge the derivatives in $p_\mu$ which can be done
by writing~\cite{Ja87,Me89} $p_\mu\longrightarrow p_\mu
+ie(\xi Q \xi^\dagger - \xi^\dagger Q \xi){\cal A}_\mu$.
Obviously this gauge term vanishes when light pseudoscalar
fields are omitted ($\xi={\rm 1\!\! I}$). Therefore such
a contribution did not arise in the work of ref.~\cite{Ja95}
as those authors were only interested in electromagnetic
processes of heavy mesons that did not contain pions.
To this end the electromagnetic current associated with the 
Lagrangian~(\ref{lheavy}) 
reads
\begin{eqnarray}
J_{\mu}^{e.m. H}&=&
ie\left(P\tilde{C}\left(D_{\mu}P\right)^{\dagger}
-D_{\mu}P\tilde{C}P^{\dagger}\right)
+ie\left(Q^{\nu}\tilde{C}Q_{\mu\nu}^{\dagger}
-Q_{\mu\nu}\tilde{C}Q^{\nu\dagger}\right)\cr
&&+ieMd\left(Q_\mu(\xi Q\xi^\dagger-\xi^\dagger Q\xi)P^\dagger
-P(\xi Q\xi^\dagger-\xi^\dagger Q\xi)Q_\mu^\dagger\right)\cr
&&+ied\epsilon_{\mu\alpha\beta\nu}
\left(Q^{\beta}p^{\nu}\tilde{C}Q^{\alpha\dagger}
-Q^{\alpha}\tilde{C}p^{\nu}Q^{\beta\dagger}\right.\cr
&&\hspace{0.5cm}\left.+\frac{i}{4}\left[
Q^{\nu\alpha}(\xi^\dagger Q\xi-\xi Q\xi^\dagger)Q^{\beta\dagger}
+Q^\beta(\xi^\dagger Q\xi-\xi Q\xi^\dagger)
Q^{\nu\alpha\dagger}\right]\right)\cr
&&+ie\frac{2\sqrt{2}c}{m_{V}}\epsilon_{\mu\alpha\beta\nu}
\left(P\tilde{C}F^{\alpha\beta}(\rho)Q^{\nu\dagger}
-Q^\nu F^{\alpha\beta}(\rho)\tilde{C}P^{\dagger}\right)
\end{eqnarray}
where $\tilde{C}=C+\frac{\alpha-1}{2}
(\xi Q\xi^\dagger+\xi^\dagger Q\xi)$ and $C$ is the charge of the 
heavy quark that is contained in the heavy meson ($C=2/3$ and $C=-1/3$ 
in the charm and bottom sectors, respectively).
Substituting both the (rotating) soliton and the bound 
state wave functions we compute
\begin{equation}
\frac{1}{2}\int d^3r 
\left(\vec{r\,}\times \vec{J\,}_{\rm e.m.}\right)_3 =
\mu_{S,0}\alpha^2\Omega_3+\mu_{V,0}D_{33}(A) 
+\mu_{S,1}\rho^\dagger\frac{\tau_3}{2}\rho
+\mu_{V,1}D_{33}(A)\rho^\dagger\rho\,.
\label{magop1}
\end{equation}
Here $\mu_{S,0},\ldots,\mu_{V,1}$ are functionals of all 
radial profiles that have been computed earlier. The functionals
$\mu_{S,0}$ and $\mu_{V,0}$ do not contain the heavy meson 
wave functions and are listed in refs.~\cite{Me89,Pa91}. Note that 
in the first term on the right hand side we made explicit the 
appearance of the moment of inertia,
$\alpha^2$ to make contact with earlier work~\cite{Ku89,Oh91}. The
detailed form of the functionals $\mu_{S,1}$ and $\mu_{V,1}$ that 
contain the bound state wave functions is presented in the appendix,
see also ref.~\cite{Sch03}. 
Using the quantization rules for the 
collective coordinates discussed above we can sandwich this
operator between baryon states,
\begin{equation}
\mu_B=\frac{1}{2}\langle B|
\int d^3r \left(\vec{r\,}\times \vec{J\,}_{\rm e.m.}\right)_3
|B\rangle
\label{magop2}
\end{equation}
yielding~\cite{Ku89,Oh91} 
\begin{eqnarray}
\mu(p)&=&\fract{1}{2}\mu_{S,0}+\fract{2}{3}\mu_{V,0} 
\qquad 
\mu(n)=\fract{1}{2}\mu_{S,0}-\fract{2}{3}\mu_{V,0} \cr
\mu(\Lambda_Q)&=&\fract{1}{2}\tilde{\mu}_{S,1}
\cr
\mu(\Sigma^{I_3=+1}_Q)&=&\fract{2}{3}\mu_{S,0}+\fract{2}{3}\mu_{V,0}
-\fract{1}{6}\tilde{\mu}_{S,1}+\fract{2}{3}\mu_{V,1}
\cr
\mu(\Sigma^{I_3=0}_Q)&=&\fract{2}{3}\mu_{S,0}
-\fract{1}{6}\tilde{\mu}_{S,1}
\cr
\mu(\Sigma^{I_3=-1}_Q)&=&\fract{2}{3}\mu_{S,0}-\fract{2}{3}\mu_{V,0}
-\fract{1}{6}\tilde{\mu}_{S,1}-\fract{2}{3}\mu_{V,1}
\cr
\mu(\Sigma^{I_3=0}_Q\to\Lambda_Q)&=&-\fract{2}{3}\mu_{V,0}
-\fract{2}{3}\mu_{V,1}
\label{magmoms}
\end{eqnarray}
where $\tilde{\mu}_{S,1}=\chi\mu_{S,0}-\mu_{S,1}$ arises
from the quantization rule 
$\alpha^2\vec{\Omega}=\vec{J}^{\rm sol}-\chi\vec{G}$.
In table~\ref{tab_mag} we present numerical results for
the magnetic moments in eq.~(\ref{magmoms}). Those results
originate from $\mu_{S,1}=-0.20 (0.05) {\rm n.m.}$ and 
$\mu_{V,1}=-0.04 (-0.01) {\rm n.m.}$ in the charm (bottom) sectors.

\begin{table}
\caption{\label{tab_mag}Preliminary results for the 
magnetic moment of baryons with a heavy quark in 
nucleon magnetons. For comparison, we predict 
$\mu(n)=2.58$ and $\mu(n)=-2.26$ which compares
reasonably well with the empirical data 
$2.79$ and $-1.91$.~\bigskip }
\begin{tabular}{c c c c c c c c c c }
$\Lambda_c$ &
$\Sigma^{++}_c$ & $\Sigma^0_c$ & $\Sigma^-_c$ &
$\Sigma^+_c\to\Lambda_c$ &
$\Lambda_b$ &
$\Sigma^0_b$ & $\Sigma^0_b$ & $\Sigma^-_b$ &
$\Sigma^+_b\to\Lambda_b$ \\
$0.12$ & $2.56$ & $0.17$ & $-2.22$ & $-2.39$&
$-0.02$ & $2.63$ & $0.21$ & $-2.19$ & $2.41$
\end{tabular}
\end{table}
We note that, in particular for $\Lambda_c$, our results compare 
reasonably well to those predicted from a spectral sum rule 
analysis~\cite{Zhu97}. On the other hand, our predictions for
the transition magnetic moments $\Sigma_Q\to\Lambda_Q$ are slightly
larger (in magnitude) than those obtained from QCD sum rules~\cite{Al01}.

We furthermore observe that with increasing masses of the heavy mesons
the parameters $\chi$, $\mu_{S,1}$ and $\mu_{V,1}$ tend to
zero. Then all magnetic moments are dominated by their 
light meson components~\cite{Oh95}. Of course, this is merely 
a reflection of the heavy quark symmetry which implies that 
the magnetic moment of the heavy quark vanishes as its mass 
becomes infinite.

\section{Conclusions}

In this talk we have presented a study on the magnetic moments
of baryons with a heavy quark. The picture for such heavy baryons
is that of a heavy meson being bound to a baryon that is built
out of light quarks. We adopt the soliton picture for baryons
in which light baryons are represented by soliton excitations
of light meson fields. This is in the spirit of the Skyrme model 
but we have also included light vector meson degrees of freedom
in the construction of the soliton. Such extensions by short
range fields are needed for various phenomenological reasons.

We have obtained the heavy meson bound state from a relativistic
Lagrangian that does not manifestly reflect the heavy quark symmetry.
Rather this model is developed for physical values of the 
heavy meson masses. However, in the (academic) limit of large
masses it is demanded to reflect the heavy quark symmetry. This 
condition strongly constrains the number of free parameters.
Also, it requires the model to contain heavy vector meson fields
rather than just the pseudoscalar degrees of freedom as 
{\it e.g.} in the bound state approach to the Skyrme model
and its application to the charm sector~\cite{Oh91}.
We have already observed in ref.~\cite{Sch95heavy} that
such extensions are also required to properly reproduce 
the spectrum of baryons with a heavy quark.

Once the model has been set up, it is straightforward to
extract the operator of the electromagnetic current by
gauging the action with respect to the photon field. In turn 
we have then used the methods to of refs.~\cite{Ku89,Oh91}
to compute the magnetic moments of baryons with a 
heavy quark. At this time the results are still somewhat
preliminary and in future we wish to further explore the connection
with the heavy quark limit for the magnetic moments~\cite{Lee00}.

We hope that Joe considers this study not only an interesting 
application of ideas that he decisively helped to develop, but
also worth to be reported on the occasion of his birthday. 


\begin{theacknowledgments}
HW would like to thank Joe Schechter for much help and valuable
advice that he provided over many years.
HW would also like to thank the organizers, in particular
A. Fariboz, for providing this worthwhile and pleasant workshop.

\noindent
This work has been supported in part by
DFG under contract no. We-1254/3--2.
\end{theacknowledgments}

\appendix
\section{Appendix}

In this appendix we list the explicit expressions for the
parameters in eq.~(\ref{magop1}) that involve the P--wave 
bound state wave--function~(\ref{pansatz}). 

For the isoscalar case we have
{\small
\begin{eqnarray}
\mu_{S,1}&=&\frac{2e}{3}\left[C+\frac{\alpha-1}{6}\right]
\int_0^\infty dr r^2 m_S(r) \cr
m_S(r)&=&-2\left(1+\fract{1}{2}R_\alpha\right)\Phi^2\cr
&&+\left(\Psi_2^\prime r+\Psi_2+R_\alpha\Psi_1\right)\Psi_1
-\left(1+\fract{1}{2}R_\alpha\right)\Psi_2^2
-\left[R_\alpha\Psi_0
+(\epsilon-\fract{\alpha}{2}\omega)\Psi_2\right]\Psi_0\cr
&&-d\left(\frac{r}{2}F^\prime\Psi_2-{\rm sin}F\Psi_1\right)\Psi_0
+\frac{2\sqrt{2}c}{gm_V}\left(r\omega^\prime\Psi_2
-2G^\prime\Psi_0\right)\Phi\,,
\label{ms1}
\end{eqnarray}
}~\hskip-0.2cm 
where again $R_\alpha={\rm cos}F-1+\alpha\left(1+G-{\rm cos}F\right)$ and
primes denote derivatives with respect to $r$. For the isovector
contribution we find
{\small
\begin{eqnarray}
\mu_{V,1}&=&\frac{e}{6} \int_0^\infty dr r^2
\left[(\alpha-1)m^{(1)}_V(r)+m^{(2)}(r)\right]\cr
m^{(1)}_V(r)&=&2\left(1+\fract{1}{2}R_\alpha\right)\Phi^2{\rm cos}F\cr
&&+\Big\{\left(\Psi_2^\prime r+\Psi_2+R_\alpha\Psi_1\right)\Psi_1
+\left(1+\fract{1}{2}R_\alpha\right)\Psi_2^2
-\left[R_\alpha\Psi_0
+(\epsilon-\fract{\alpha}{2}\omega)\Psi_2\right]\Psi_0\Big\}
{\rm cos}F\cr
&&+d\,{\rm sin}2F\, \Psi_0\Psi_1
-\frac{2\sqrt{2}c}{gm_V}\left(r\omega^\prime\Psi_2
-2G^\prime\Psi_0\right)\Phi\,{\rm cos}F\, , \cr
m^{(2)}_V(r)&=&-2Md\,r\,{\rm sin}F\, \Psi_2\Phi\cr
&&+d\left[r\left(\Psi_0\Psi_2^\prime-\Psi_0^\prime\Psi_2\right)
+\Psi_0\Psi_2+2R_\alpha\Psi_0\Psi_1
+2r(\epsilon-\fract{\alpha}{2})\Psi_1\Psi_2\right]
\label{mv1}
\end{eqnarray}
}~\hskip-0.1cm 
The analytic expressions for $\mu_{S,0}$ and $\mu_{V,0}$ can
be found in refs.~\cite{Me89,Pa91}.


\end{document}